# Atomic-Scale Probing of Heterointerface Phonon Bridges in Nitride Semiconductor

Yue-Hui Li[1,2], Rui-Shi Qi[1,2], Ruo-Chen Shi[1,2], Jian-Nan Hu[3], Zhe-Tong Liu[2], Yuan-Wei Sun[1,2], Ming-Qiang Li[2], Ning Li[2], Can-Li Song[3], Lai Wang[4], Zhi-Biao Hao[4], Yi Luo[4], Qi-Kun Xue[3,5,6], Xu-Cun Ma[3], Peng Gao[1,2,7,8]*

[1]International Center for Quantum Materials, School of Physics, Peking University, Beijing, 100871, China
[2]Electron Microscopy Laboratory, School of Physics, Peking University, Beijing, 100871, China
[3]State Key Laboratory of Low-Dimensional Quantum Physics and Department of Physics, Tsinghua University, Beijing 100084, China
[4]Beijing National Research Center for Information Science and Technology and Department of Electronic Engineering, Tsinghua University, Beijing 100084, China
[5]Beijing Academy of Quantum Information Sciences, Beijing 100193, China
[6]Southern University of Science and Technology, Shenzhen 518055, China
[7]Interdisciplinary Institute of Light-Element Quantum Materials and Research Center for Light-Element Advanced Materials, Peking University, Beijing 100871, China.
[8]Collaborative Innovation Centre of Quantum Matter, Beijing 100871, China

*Corresponding author. E-mail: p-gao@pku.edu.cn

**Interface phonon modes that are generated by several atomic layers at the heterointerface play a major role in the interface thermal conductance for nanoscale high-power devices such as nitride-based high-electron-mobility transistors and light emitting diodes. Here we measure the local phonon spectra across AlN/Si and AlN/Al interfaces using atomically resolved vibrational electron energy-loss spectroscopy in a scanning transmission electron microscope. At the AlN/Si interface, we observe various localized phonon modes, of which the extended and interfacial modes act as bridges to connect the bulk AlN modes and bulk Si modes, and are expected to boost the inelastic phonon transport thus substantially contribute to interface thermal conductance. In comparison, no such phonon bridge is observed at the AlN/Al interface, for which partially extended modes dominate the interface thermal conductivity. This work provides valuable insights into understanding the interfacial thermal transport in nitride semiconductors and useful guidance for thermal management via interface engineering.**

Rapid developments of various modern information technologies such as big data transmission, cloud computing, artificial intelligence technology, and the Internet of things, have put forward higher requirement on network transmission speed and capacity, demanding for higher-power and higher-speed electronic devices [1,2] such as nitride-based high-electron-mobility transistors (HEMTs) [3]. Thermal management in such devices becomes crucial as the high output power density results in strong Joule self-heating effect, which increases the channel temperature and severely degrades device performance [4,5]. Solutions for thermal management include searching for high thermal conductivity materials [6,7] and increasing interface thermal conductance (ITC) via interface engineering [8–10]. The latter approach becomes increasingly important when the size of the device approaches nanoscale as the ITC dominates the device thermal resistance [11,12]. However, it is challenging to obtain the precise knowledge of ITC due to the atomic size and the buried nature of heterointerfaces. The common methods to characterize thermal conductivity, including the time-domain thermoreflectance [13], the frequency-domain thermoreflectance [14], the 3-ω method [15] and the coherent optical thermometry [16], suffer from a poor spatial resolution that is insufficient to measure thermal properties at the nanoscale.

In fact, the thermal properties of semiconductor and insulator interfaces are largely governed by the interface phonons, and interface phonons also dominate the ITC of metal/semiconductor interfaces because electron-phonon coupling has little effect on the ITC of metal/semiconductor interfaces and can be ignored [17,18]. Previous calculations indicate that the interface can bridge the phonons with different energies and thus boost the inelastic phonon transport [19]. Recent methods such as modal analysis were used to correlate the interface phonons with interfacial heat flow [20–23].

Specifically, the interface phonon modes can be classified into four classes: extended, partially extended, isolated and interfacial modes, based on how the vibrational energy is distributed in space [22]. The atomic vibrations are delocalized into both sides of the interface for extended modes and are localized on one side for partially extended modes. Isolated modes are not present at the interface, while interfacial modes are highly localized at the atomically thin interface. Indeed, the interfacial modes usually have the highest contribution to the ITC on a per mode basis [19,22,24,25]. Very recently, the interfacial modes for $MoS_2/WSe_2$ [26], Ge/Si [27], $SrTiO_3/CaTiO_3$ superlattices [28] and cubic-BN/diamond system [29] have been experimentally observed. At a Ge/Si interface, calculations suggested a small number of interfacial modes make substantial contributions to ITC acting as bridges to connect the bulk modes of two sides, while the isolated modes hardly contribute to ITC [30]. However, the dominant type of interface phonon for ITC is likely different in different material systems depending on the interface bonding. It is thus useful to experimentally study nanoscale phonon behaviors at various interfaces and correlate them to thermal conductance across the interface.

In this work, we study interface phonons at the AlN/Si and AlN/Al heterointerfaces. Due to the wide bandgap, high-breakdown electric field, and high carrier mobility, the III-V nitride semiconductors such as GaN, AlN and their ternary compounds, are considered promising in the next generation high-power and high-frequency electronic devices [31], which, however, requires excellent thermal conductivity, especially high ITC [32]. For these nitride heterostructures, it remains largely unknown what types of interface phonons exist, not to mention how they impact on the ITC and which one dominates. By using the vibrational electron energy loss spectroscopy (EELS) in scanning transmission electron microscope (STEM) that have the abilities to measure the phonon spectra [33–36] at atomic scale [26,27,37–41], we probe the interface phonons to gain insights into interface thermal properties.

We observe multiple types of interface phonons at an AlN/Si interface and find that the extended modes that connect the bulk AlN phonon modes and bulk Si phonon modes, and the interfacial mode that promotes transverse acoustic (TA) mode of Si to penetrate into AlN layer, act as bridges to mainly contribute to ITC. However, no such bridging effect is observed across AlN/Al interface where the partially extended modes dominate, while the extended modes and interfacial modes only take up a small proportion in the total modes. AlN/Si interface is therefore expected to have a much higher ITC than AlN/Al interface since the extended modes contribute 2-3 times larger than partially extended modes to the ITC according to the molecular dynamics (MD) simulations. Our results unveil the very different interface phonon modes at the heterointerfaces of AlN/Si and AlN/Al, and find they have very different contributions to the ITC, providing new insights into understanding and engineering the interface thermal properties.

The AlN (0001) film was grown on Si (111) substrate. To exclude the long-range dipole scattering and significantly improve the relative contribution of highly localized impact scattering signal, the direct beam is moved off the optical axis and outside the EELS aperture, which enables us to obtain the local vibrational information [41]. Figure 1(a) displays the atomic-resolution high angle annular dark field (HAADF) image of AlN/Si interface along the $[11\bar{2}0]$ direction of AlN and the $[1\bar{1}0]$ direction of Si, wherein the Si and Al atoms form covalent Si-Al bonds and would significantly influence the vibrational properties near the interface.

Figure 1(b) displays the EEL spectral mapping across the AlN/Si interface. The lineprofile of EEL spectra are provided in Fig. S1. In the bulk regions, the characteristic vibrational peaks can be assigned based on the MD calculations and refs [42–44]. For the Si bulk phonon spectrum, three peaks at 19.8 meV, 43.8 meV and 59.0 meV are assigned to Si-TA1 mode, longitudinal acoustic/longitudinal optical (Si-LA1/LO1) mode and transverse optical (Si-TO1) mode, respectively. For the AlN bulk phonon spectrum, four peaks at 31.5 meV, 45.0 meV, 63.2 meV and 80.0 meV are AlN-TA2 mode, AlN-LA2/TO2 mode, AlN-TO3 mode and AlN-LOs/TOs mode, respectively. It should be noted that AlN-LOs/TOs mode is composed of several optical phonon branches with close frequencies. We find that Si-TA1 mode penetrates into AlN layer, Si-LA1/LO1 mode connect to AlN-TA2 mode, and Si-TO1 mode also connects to AlN-TO3 mode at the interface forming bridges across the interface. AlN-LOs/TOs mode penetrate into Si layer but decays rapidly within ~1 nm.

In order to gain insights into the physical origin of different spectral features at the interface, MD simulation was performed using the Large-scale Atomic/Molecular Massively Parallel Simulator (LAMMPS) package [45,46] with the Stillinger-Weber potential [47,48]. Figure 2(a) and 2(b) show the measured phonon spectra and the calculated projected phonon density of states (PPDOS) of bulk AlN, AlN/Si interface and bulk Si. The arrows represent the energy shift of Si-TA1 mode on going from Si layer to the interface and AlN-TA2 mode on going from AlN layer to the interface, respectively. The experiments in Fig. 2(a) show that Si-TA1 mode has a blueshift of ~3.4 meV and AlN-TA2 mode has a blueshift of ~1.5 meV at the interface, which is in reasonable agreement with the simulated spectra in Fig. 2(b) with a blueshift of ~4.8 meV for Si-TA1 mode and ~1.2 meV for AlN-TA2 mode.

To quantitatively validate these spectral features, the background-subtracted vibrational spectra were fitted using a simple Gaussian peaks fitting model. Figure 2(c) and 2(d) show the fitted peak positions and peak intensities across the AlN/Si interface respectively. Si-TA1 mode penetrates into AlN layer and has a slight blueshift with intensity decreased by ~15 times within 1 nm. Si-LA1/LO1 mode and AlN-TA2 mode have an energy shift near the interface and connect to each other to form a phonon bridge. Si-TO1 mode and AlN-TO3 mode also have a slight energy shift and connect to the other as a phonon bridge. These phonon bridges connect the phonons with significantly different energies and promote phonon transport across the interface to contribute thermal conductance [19]. AlN-LA2/TO2 mode just exists in AlN layer and has a sharp intensity decrease near the interface, which contributes little to thermal conductance. AlN-LOs/TOs mode penetrates into Si layer and also makes a contribution to ITC.

Figure 2(e) show energy-filtered EELS maps of five typical phonon branches in Fig. 2(c). Figure 2(f) shows the corresponding calculated vibrational eigen vector displacements. In the intensity map of 24-26 meV, the vibrational intensity is localized near the interface with a peak width of 1.5-2 nm, and the simulated vibrational eigen vector displacement below also displays that the vibration is localized at the interface, indicating an interfacial mode. The intensity maps of 32-34 meV and 61-63 meV showing strong intensities in both sides, are extended modes as further confirmed by the simulated eigen vector displacements. The intensity map of 45-47 meV corresponds to an isolated mode which is not present at the interface, while the intensity map of 84-86 meV corresponds to partially extended mode whose vibration is mainly localized in the AlN side.

In comparison, we also measured the phonon spectra across an AlN/Al interface as an example of typical semiconductor/metal-electrode interface. Figure 3(a) displays the atomic-resolution HAADF image and atomic structure of AlN/Al interface, where the Al atoms on the Al (111) plane and the Al atoms on AlN (0001) plane form Al-Al metallic bonds like those in bulk Al layer. Figure 3(b) shows mapping of the background-subtracted vibrational spectra across AlN/Al interface, and Figure 3(c) displays the typical EEL spectra of bulk AlN, bulk Al and AlN/Al interface. The lineprofile of EEL spectra are provided in Fig. S2. Different from AlN/Si interface, no distinct phonon bridges are observed to connect the phonons with significantly different energies in two sides. The fitted peak position line profile in Fig. 3(d) shows Al-TA3 mode and Al-LA3 mode terminate at the interface with a slight energy shift. AlN-TO3 mode just abruptly disappeared at the interface, while the AlN-LOs/TOs mode penetrates into Al layer for ~1.5 nm. Most of these modes are partially extended modes, as the intensity maps shown in Fig. S3.

Table I shows the calculated modal contribution to ITC (denoted by G) for each of mode class at AlN/Si and AlN/Al interfaces. The extended modes have vibration components on both sides, and act as bridges for the phonons of the two sides and have more contribution to ITC on a per mode basis (larger G/DOS) than partially extended modes. In fact, the extended modes comprise ~45% of total DOS in AlN/Si system, but only ~22% of total DOS in AlN/Al system, which explains that the ITC of AlN/Si ($\sim 300$ $MW \cdot m^{-2} \cdot K^{-1}$) [49] is larger than that of AlN/Al ($\sim 117$ $MW \cdot m^{-2} \cdot K^{-1}$) [50]. Our non-equilibrium MD simulation also obtain the same conclusion, i.e., the ITC of AlN/Si ($\sim 330$ $MW \cdot m^{-2} \cdot K^{-1}$) is higher than that of AlN/Al ($\sim 103$ $MW \cdot m^{-2} \cdot K^{-1}$). Isolated modes contribute little to ITC because there is no vibration near the interface [22]. Although both our results and ref [22] show that interfacial modes have the highest contribution (G/DOS) on a per mode basis, their low population makes them contribute no more than 10% of total ITC.

In summary, we have directly measured the interface phonon modes of AlN/Si and AlN/Al heterointerfaces at atomic-scale using STEM-EELS. We experimentally obtain and distinguish different interface phonon modes at the AlN/Si system, and find that the extended modes and interfacial modes can act as phonon bridges to contribute heavily to the ITC. In contrast, no obvious phonon bridge is observed in AlN/Al system since it has low population for extended modes but very high population for partially extended modes that contribute less to ITC. These features account for their significantly different ITC. The demonstrated ability to directly correlate the measured interface phonons with thermal transport properties at heterointerfaces on an atomic level opens new avenues for engineering the interface thermal properties for thermal management and thermoelectric devices.

**Acknowledgements**

The work was supported by the National Key R&D Program of China (2019YFA0708202), the National Natural Science Foundation of China (11974023, 52021006, 62004113, 61991443), Key-Area Research and Development Program of Guangdong Province (2018B030327001, 2018B010109009), the "2011 Program" from the Peking-Tsinghua-IOP Collaborative Innovation Center of Quantum Matter, Youth Innovation Promotion Association, CAS. We acknowledge Electron Microscopy Laboratory of Peking University for the use of electron microscopes. We acknowledge High-performance Computing Platform of Peking University for providing computational resources for the MD calculation.

**Author contributions**

P.G. and Y.L. conceived the project. Y.L. designed and performed the EELS measurements. R.Q. and R.S. wrote the data processing codes, and N.L. helped the data analysis. Y.L. performed the MD simulation. J.H. grew the sample under the direction of C.S., X.M., Q.X., L.W., Z.H. and Y.L. Y.L and Z.L. prepared the TEM samples. Y.S. and M.L. acquired the atomic-resolution HAADF images. P.G. supervised the project. All authors discussed the results and commented on the manuscript.

**Competing interests**

The authors declare no competing interests.

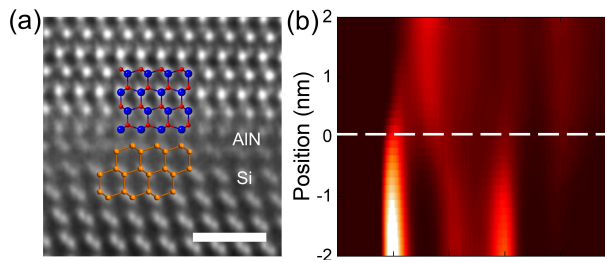

FIG. 1. Atomic structure and phonon spectra of AlN/Si interface. (a) Atomic HAADF images and structure of AlN/Si interface. Scale bar, 1 nm. (b) Mapping of phonon EEL spectra across the AlN/Si interface.

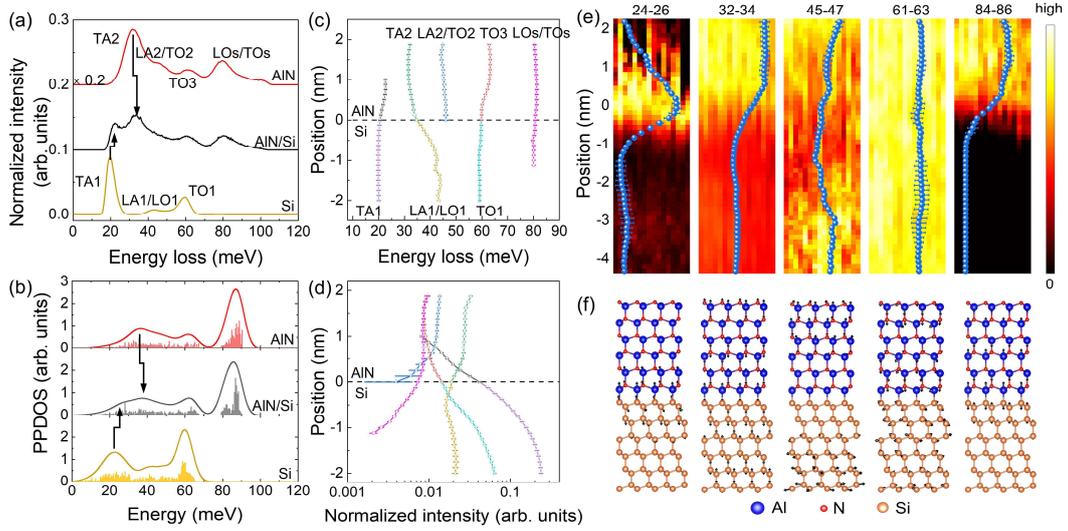

FIG. 2. Interface phonons at the AlN/Si interface. (a) Measured EEL spectra and (b) calculated PPDOS of bulk AlN, AlN/Si interface, and bulk Si. The arrows represent the blueshift of Si-TA1 mode and AlN-TA2 mode at the interface. (c) Fitted phonon peak positions across AlN/Si interface. (d) Phonon peak intensities across AlN/Si interface. The peak positions and intensities are extracted by Gaussian peaks fitting method. (e) Intensity mapping of different phonon modes and the corresponding intensity line profile. (f) Corresponding calculated eigen vector displacements of interface modes.

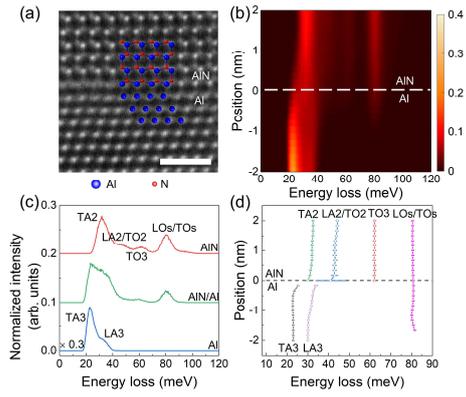

FIG. 3. Atomic structure and interface phonons of AlN/Al. (a) Atomic HAADF images and structure of AlN/Al interface. Scale bar, 1 nm. (b) Mapping of phonon EEL spectra at AlN/Al interface. (c) EEL spectra of bulk AlN, AlN/Al interface, and bulk Al. (d) Phonon peak positions across AlN/Al interface.

| Mode Type | | DOS(%) | G(%) | G/DOS |
|---|---|---|---|---|
| Extended | AlN/Si | 44.93 | 66.96 | 1.49 |
| | AlN/Al | 14.07 | 22.36 | 1.59 |
| Partially extended | AlN/Si | 54.03 | 27.95 | 0.52 |
| | AlN/Al | 83.37 | 67.49 | 0.81 |
| Isolated | AlN/Si | 0.16 | 0.005 | 0.03 |
| | AlN/Al | 1.11 | 0.01 | 0.01 |
| Interfacial | AlN/Si | 0.88 | 5.09 | 5.78 |
| | AlN/Al | 1.45 | 10.14 | 6.99 |

Table. I. Density of states and contribution to ITC for the four different classes of vibration across AlN/Si interface and AlN/Al interface. Column 2 represents the fraction of total DOS. Column 3 represents the fraction of total ITC. Column 4 represents the contribution to ITC per mode.

# Supplemental Material: Atomic-Scale Probing of Heterointerface Phonon Bridges in Nitride Semiconductor

Yue-Hui Li[1,2], Rui-Shi Qi[1,2], Ruo-Chen Shi[1,2], Jian-Nan Hu[3], Zhe-Tong Liu[2], Yuan-Wei Sun[1,2], Ming-Qiang Li[2], Ning Li[2], Can-Li Song[3], Lai Wang[4], Zhi-Biao Hao[4], Yi Luo[4], Qi-Kun Xue[3,5,6], Xu-Cun Ma[3], Peng Gao[1,2,7,8]*

[1]International Center for Quantum Materials, School of Physics, Peking University, Beijing, 100871, China
[2]Electron Microscopy Laboratory, School of Physics, Peking University, Beijing, 100871, China
[3]State Key Laboratory of Low-Dimensional Quantum Physics and Department of Physics, Tsinghua University, Beijing 100084, China
[4]Beijing National Research Center for Information Science and Technology and Department of Electronic Engineering, Tsinghua University, Beijing 100084, China
[5]Beijing Academy of Quantum Information Sciences, Beijing 100193, China
[6]Southern University of Science and Technology, Shenzhen 518055, China
[7]Interdisciplinary Institute of Light-Element Quantum Materials and Research Center for Light-Element Advanced Materials, Peking University, Beijing 100871, China.
[8]Collaborative Innovation Centre of Quantum Matter, Beijing 100871, China

*Corresponding author. E-mail: p-gao@pku.edu.cn

**Sample fabrication.** In a molecular beam epitaxy chamber, the Al/AlN sample was grown on Si(111) substrate. As the growth procedure reported before [1], substrate nitridation for 60 s and alumination under Al flux irradiation for 900 s were performed for formation of AlN nucleation layer on Si(111) with sharp interface. After the AlN main layer growth, the nitrogen source was turned off, and substrate temperature was lowered for nominally 40 nm Al metal layer growth. Reflection high-energy electron diffraction (RHEED) patterns of Al layer were kept streaky, and Kikuchi lines were observed, indicating Al layer is with two-dimensional growth mode and flat surface. Finally, $AlO_x$ were formed on the top by natural oxidation after exposure to atmosphere for protection of Al layer.

**TEM sample preparation.** The cross-sectional TEM specimen was prepared by conventional mechanical polishing followed by an argon ion milling. The argon ion milling was carried out using the Precision Ion Polishing System (Model 691, Gatan Inc.). The accelerating voltage was first set 3.5 kV with angles ± 6º until a hole was made, and then lowered to 1.0 kV with angles ± 3º until the hole was enlarged to the interface, and finally lowered to 0.1 kV for 2 min to remove the surface amorphous layer with free damage. The specimen was baked at 160º for 16 hours to further remove the surface amorphous layer before EELS experiments.

**Experimental setup.** The vibrational spectra were acquired at a Nion U-HERMES200 electron microscope equipped with both the monochromator and the aberration corrector operated at 30 kV. The probe convergence semi-angle and the collection semi-angle were both 25 mrad. The electron beam was moved off optical axis with 80 mrad for off-axis experiments to greatly reduce the contribution of the dipole scattering. Moreover, the vibrational spectra with such large convergence semi-angle, collection semi-angle and off-axis angle are close to the localized PDOS [2]. The energy dispersion channel was set as 0.35 meV with 2048 channels in total. The typical beam current was ~12 pA. The best energy resolution (full width of half maximum, FWHM) was 6.2 meV for on-axis experimental setup and 7.5 meV for off-axis experimental setup when the beam was focused on the specimen. The spatially resolved EEL spectra in Fig. 1(b) and Fig. 3(b) were originally recorded as a spectrum image, with single exposures of 1600 ms per pixel. The spectrum image acquired was 2 nm x 10 nm with 0.125 nm per pixel in Fig. 1(b), and 2.7 nm x 12 nm with 0.166 nm per pixel in Fig. 3(b). The HAADF image were recorded by an aberration-corrected FEI Titan Themis G2 operated at 300 kV with spatial resolutions up to 60 pm.

**EELS data processing.** All acquired vibrational spectra were processed using the Gatan Microscopy Suite and custom-written MATALAB code. The EEL spectra were first aligned by their normalized cross correlation. Next, the block-matching and 3D filtering (BM3D) algorithm were applied to remove Gaussian noise. All spectra were then normalized to the intensity of zero-loss-peak. The Background arising from both the tail of ZLP and non-characteristic phonon losses was fitted using the modified Peason-VII function [3,4] with two fitting windows and then subtracted to obtain the vibrational signal. The Lucy-Richardson algorithm was then employed to ameliorate the broadening effect caused by finite energy resolution, taking the elastic ZLP as the point spread function. The spectra were summed along the direction parallel to the interface to obtain a line-scan data with a good signal-to-noise ratio. Then the spectra within ±2 nm near the interface were cropped because the bridging effects are localized within ±1 nm near the interface. The vibrational spectra were fitted using a simple Gaussian peaks fitting model to extract the peak positions and peak intensities.

**MD simulation.** The MD simulations were performed using the Large-scale Atomic/Molecular Massively Parallel Simulator (LAMMPS) package [5] with the Stillinger-Weber (SW) potential [6,7]. The atomic plane of Si (111) and AlN (0001) are mismatched, e.g., the lattice constants are $a_{Si(111)}$ = 3.88 Å and $a_{AlN(0001)}$ = 3.15 Å experimentally. MD simulations are hard to simulate such mismatched interface since it requires a large cross-section area to fit $a_{Si(111)}$ and $a_{AlN(0001)}$. To overcome this difficulty, we can either set the average lattice constant in the directions parallel to the interface, $a_{/\!/}$ = ($a_{Si(111)}$ + $a_{AlN(0001)}$)/2 and set the lattice constant in the directions perpendicular to the interface, $a_\perp$, to give a zero stress in that direction [8], or either change the interatomic potential of Si to be the same as AlN so that they have the same lattice constant $a_{AlN(0001)}$ [9]. We choose the second method for it works better near the interface [9]. The computational code, Dynaphophy, is used to extract the phonon properties from MD simulations [10].

Recently developed interface conductance modal analysis (ICMA) [11,12] enables us to determine the modal contribution to ITC by substituting the modal contributions into either an equilibrium MD (EMD) or non-equilibrium MD (NEMD) expression for ITC. The modal ITC is

$$G_n = \frac{1}{k_B A T^2} \int \langle J_n(t) J(0) \rangle dt$$

Where $k_B$ is the Boltzmann constant, $A$ is the cross-sectional contact area, $T$ is the absolute temperature. $J_n(t)$ and $J(0)$ are the mode-specific heat current with a separation time of $t$. The $\langle ... \rangle$ brackets indicate averaging over different time origins. Based on the magnitude of the eigen vectors [11], the modes were classified into four distinct classes: extended, partially extended, isolated and interfacial modes, and then we can obtain their contribution to interface thermal conductance.

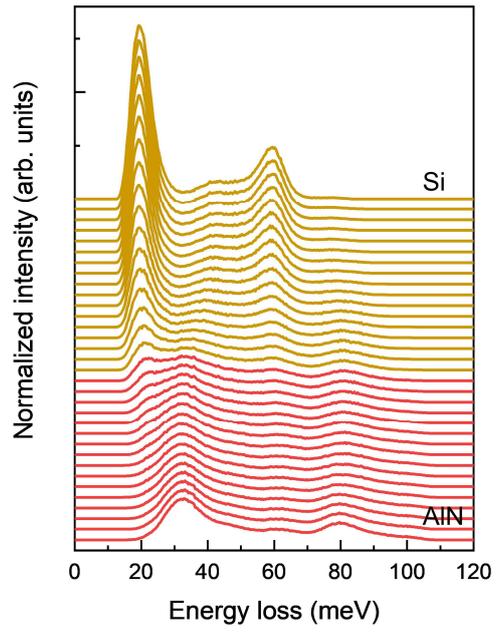

Fig. S1. Lineprofile of localized vibrational signal across AlN/Si interface with 0.125 nm every stack.

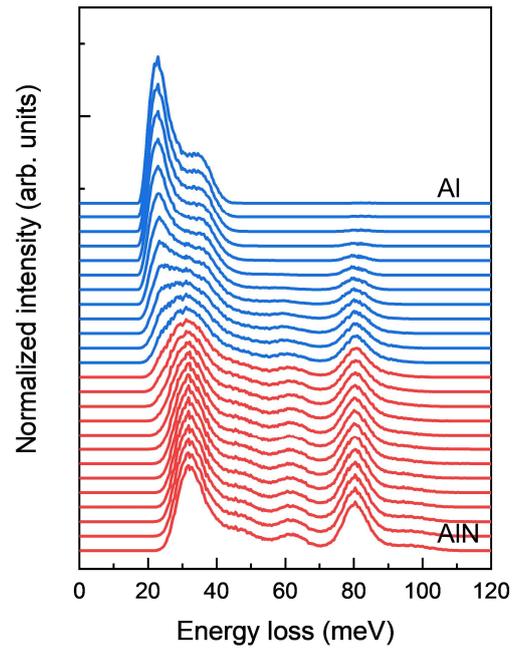

Fig. S2. Lineprofile of localized vibrational signal across AlN/Al interface with 0.166 nm every stack.

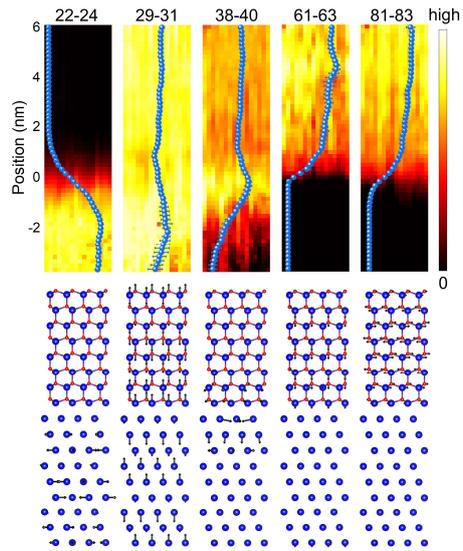

Fig. S3. Mapping of EELS signal of different phonon modes and corresponding lineprofile of signal across AlN/Al interface. The panel below shows examples of eigen vector displacements.